\documentclass{iopart}
\usepackage{graphicx}
\usepackage[latin1]{inputenc}

\newcommand{\ket}[1]{|#1\rangle}

\newcommand{\bk}[2]{\langle#1|#2\rangle}
\newcommand{\eq}[1]{Eq.~(\ref{#1})}

\newcommand{\fig}[1]{Fig.~(\ref{#1})}
\newcommand{\virg}[1]{``#1''}

\begin{document}

\title{Engineering massive quantum memories by
topologically time-modulated spin rings}

\author{S. M. Giampaolo, F. Illuminati, A. Di Lisi, S. De Siena}
\address{Dipartimento di Fisica ``E. R. Caianiello'', Universit\`a
di Salerno, INFM UdR di Salerno, INFN Sezione di Napoli,Gruppo
Collegato di Salerno, Via S. Allende, 84081 Baronissi (SA), Italy}

\begin{abstract}
We introduce a general scheme to realize perfect storage of
quantum information in systems of interacting qubits. This novel
approach is based on {\it global} external controls of the
Hamiltonian, that yield time-periodic inversions in the dynamical
evolution, allowing a perfect periodic quantum state
recontruction. We illustrate the method in the particularly
interesting and simple case of spin systems affected
by $XY$ residual interactions with or without static
imperfections. The global control is achieved by step
time-inversions of an overall topological phase of the
Aharonov-Bohm type. Such a scheme holds both at finite size and in
the thermodynamic limit, thus enabling the massive storage of
arbitrarily large numbers of local states, and is stable against
several realistic sources of noise and imperfections.
\end{abstract}


\submitto{\JPB}

\maketitle

\section{Introduction}

In the attempt to pave the way to the realization of scalable
schemes for quantum computation, much theoretical work has been
recently aimed at developing suitable strategies for the efficient
processing and the coherent transfer of quantum information
\cite{Bose,Depasquale,Subramanyam,Datta,Song}. Besides these two
fundamental aspects, a further crucial requirement for the
realization of scalable quantum computers is the possibility to
store quantum data on time scales at least comparable to those
needed for the computational process. In particular, it is very
important to introduce systems acting as stable and robust quantum
memories that recover and conserve large sets of quantum states
that would be otherwise usually lost in very short times, due to
quantum diffusion and decoherence \cite{Song}.

To ensure stable information storage in a quantum register, many
different noise-evading schemes have been proposed
\cite{Song,Lukin,Lukin2,Lukin3,nano1,nano2,Zoller,Poggio,Gingrich,Kitaev}.
All these works can be roughly classified in two different groups:
the first includes schemes based on some error correction
technique. The remaining proposals exploit some intrinsic property
of the quantum register that leaves some specific subsets of
quantum states unaffected along the temporal evolution
(decoherence-free subspace schemes). These latter approaches
provide, in principle, the complete solution to the problem of
quantum information storage, but, unfortunately, they are
extremely sensitive to almost any source of imperfection. On the
contrary, the schemes based on quantum error correction techniques
are characterized by a dynamics that allows at any time the
unambiguous reconstruction of the initial information. The main
trouble with error correction techniques lies in the fact that
usually only very few states can be effectively accessed to store
information and, hence, relatively large arrays of qubits are
needed to memorize relatively small amounts of information.

In the present paper we introduce a new approach to quantum state
storage based on the idea of time-controlled periodic dynamics
that allows a perfect, periodic reconstruction of a generic
initial state. To demonstrate and describe it, let us first recall
that an ideal quantum register can be considered as a set of
isolated identical qubits subject to a local Hamiltonian
\begin{equation}
\hat{H}_0 \, = \, B\sum_{i}\sigma_{i}^{z} \; ,
\end{equation}
where $B$ is the half
gap between the two energy levels of each spin\cite{Georgeot}.
However, in realistic situations, the register is subject to noise
caused by disorder in the local gap and by interactions both with
the substrate environment and between the qubits. Typically, at
least at sufficiently low temperatures or sufficiently weak
coupling with the background substrate, the interaction with the
external environment takes place on time scales much slower than
those associated to the computational process. For instance,
in the case of quantum gate operations with hyperfine levels 
in trapped ions, the ratio $t_{gate}/t_{decoh}$ can be
as small as $10^{-9}$ \cite{Roadmap,Divincenzo}. Then, the
corrupting effects that take place on the same time scales of the
computational processes, and thus need to be addressed first, 
are those due to the unavoidable presence of the residual, deterministic
and/or random inter-qubit interactions \cite{Roadmap,Divincenzo}. These
interactions cause, in general, fast quantum diffusion and, as a
consequence, the complete corruption of the information one wishes to store
and process. Therefore, even before considering the effects of thermal
and environmental decoherence, realistic quantum registers in the presence
of irreducible noise and imperfections must be described by
Hamiltonians of the form \cite{Georgeot,Montangero}:
\begin{equation}
\label{GenericHamiltonian} \hat{H}_{tot} \, = \, \hat{H}_0 +
\hat{H}_{err} \; ,
\end{equation}
where $\hat{H}_{err}$ is the residual inter-qubit
Hamiltonian. In several physical situations, it
can be described by $XY$ interaction terms:
\begin{equation}
\label{ErrorHamiltonian} \hat{H}_{err} \, = \,
-\sum_{i}(\lambda +\eta_i)\bigg(
\sigma_{i}^{+}\sigma_{i+1}^{-} + H. c. \bigg) \; ,
\end{equation}
where the random variables of vanishing mean $\eta_i$ are the
local imperfections in the global, averaged nearest-neighbor
coupling amplitude $\lambda$, and $\sigma_{k}^{\pm} =
\sigma_{k}^{x} \pm i \sigma_{k}^{y}$. Such a site-dependent $XY$
model applies immediately to spin-$1/2$ based quantum registers
(such as in NMR devices), but gives as well an effective
description of a register based on hopping and/or interacting
particles on a lattice, in the presence of an energy gap such that
only two local states on each lattice site can be considered
\cite{Schoen}, so that the material particles are mapped in
spin-$1/2$ systems. The storing scheme that we shall describe in the
present work holds in general for many classes of inter-qubit interaction
Hamiltonians, for instance $XXZ$ interactions, but in a more limited 
range of validity when dealing with imperfections \cite{ControlJOB}.
Therefore, in the following, we will restrict our analysis to
physical systems and regimes such that the residual inter-qubit
couplings can be well described by $XY$ interaction Hamiltonians,
including local imperfections.

As already mentioned, the presence of $\hat{H}_{err}$, even if
$\lambda \ll B$ and $\eta_i =0$ $\forall \; i$, rapidly destroys
the storing capacity of the register: immediately after having
stored an {\em initial state} $\ket{\psi(t=0)} \equiv \ket{\chi}$, such
state starts to evolve and diffuse indefinitely. To illustrate and
measure this effect, we can follow the behaviour of the time-dependent
two-dimensional overlap  ${\cal{F}}_{d}(t) \equiv |\langle \chi |\psi(t)\rangle|^2$
for different stored initial states and physical situations \cite{Fazio}. 
Let us then consider a closed (periodic boundary conditions), unmodulated 
chain with $XY$ nearest neighbour residual interactions between the
spins, and let us introduce the set of one-magnon, locally excited 
states 
\begin{equation}
\ket{\Psi_d}=\ket{\uparrow_{d}}\prod_{i\neq d}\ket{\downarrow_i} \; ,
\end{equation}
where the product involves all the qubits
of the register except the one at site $d$. In Fig. 1 we take
the initial state $\ket{\chi}$ to be $\ket{\Psi_0}$, i.e. 
the one-magnon excitation placed on site $d=0$, while in Fig. 2
the initial state is taken to be the linear combination with
equal weights of one-magnon excitations placed on sites 
$d=1$ and $d=-1$.
\begin{figure}[t!]
\includegraphics[width=7.5cm]{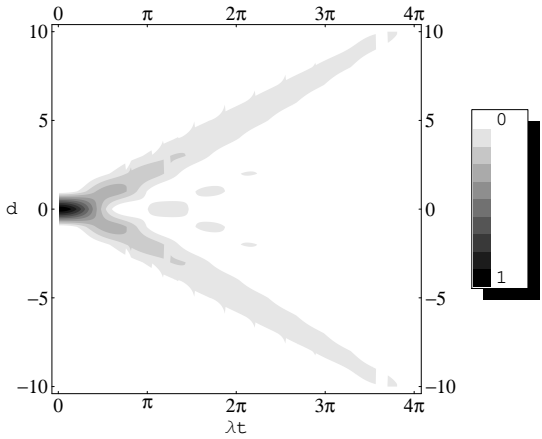}
\caption{Unmodulated register: contour plot showing the evolution
of the bidimensional overlap ${\cal{F}}_{d}(t)$, starting from the initial state
$\ket{\Psi_0}$, as a function of the distance $d$ from site $0$
($y$-axis) and of the dimensionless time $\lambda t$ in atomic
units $\hbar=1$ ($x$-axis). Here $B=100 \lambda$. The value of the
overlap increase from white $({\cal{F}}_{d}(t)=0)$ to black
$({\cal{F}}_{d}(t)=1)$.} 
\label{figura1a}
\end{figure}
\begin{figure}[t!]
\includegraphics[width=7.5cm]{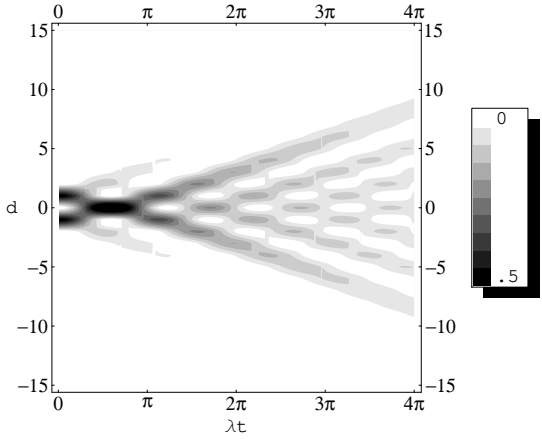}
\caption{Unmodulated register: contour plot showing the evolution
of the two-dimensional overlap ${\cal{F}}_{d}(t)$, for the initial state
$\ket{\chi} = (\ket{\Psi_1} + \ket{\Psi_{-1}})/\sqrt{2}$, as a
function of the distance $d$ from site $0$ ($y$-axis) and of the
dimensionless time $\lambda t$ ($x$-axis) in atomic unit
$\hbar=1$. The value of the overlap increases from white
(${\cal{F}}_{d}(t) = 0$) to black (${\cal{F}}_{d}(t) = 0.5$).}
\label{figura1b}
\end{figure}
By looking both at \fig{figura1a} and \fig{figura1b}, we see that,
irrespective of the different set of physical parameter and
stored states one considers, the quantum state diffusion grows 
indefinitely in time and the initial state is never recovered 
at any site of the lattice. Consequently, the one-dimensional 
fidelity of the initially stored state at a fixed position $d$ 
in the lattice quickly decrease, undergoing partial revivals, 
as shown in \fig{figura1c}. It is worth noticing that in a
non modulated register the fidelity of an initial superposition
is more robust against quantum diffusion. This intrinsic property 
is helpful in the case of modulated (controlled) registers.
\begin{figure}[t!]
\includegraphics[width=7.5cm]{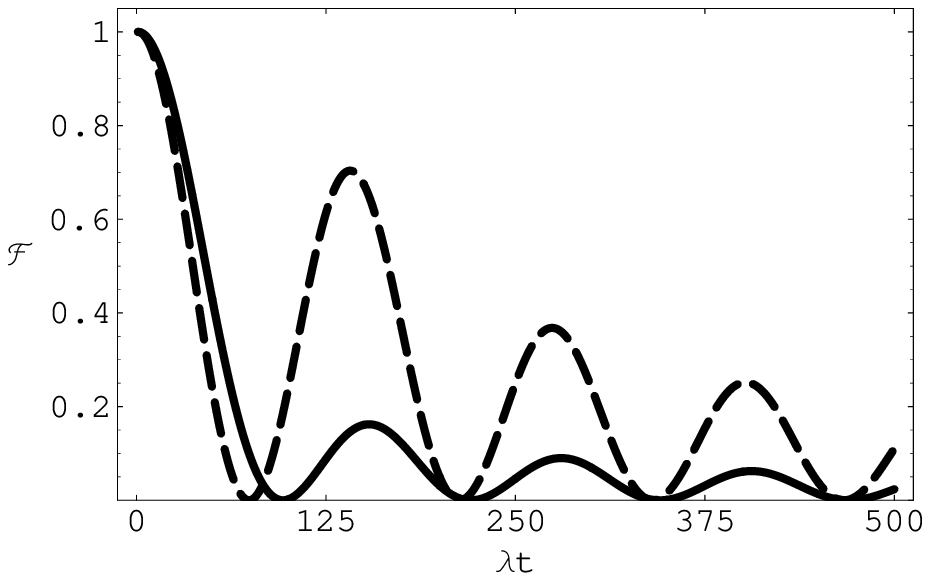}
\caption{Unmodulated register: One-dimensional fidelity 
${\cal{F}}_{0}(t)$ at fixed
lattice site $d=0$, as a function of the
dimensionless time $\lambda t$ for different initially 
stored states. The solid line represents the evolution
of the fidelity with initial state taken to be the single-magnon
excitation $\ket{\Psi_0}$ localized at site $d=0$.
The dashed line represents the evolution of
the one-dimensional fidelity for the initial superposition
state $\ket{\chi} = (\ket{\Psi_1} + \ket{\Psi_{-1}})/\sqrt{2}$.
Notice that superpositions are more robust against quantum
diffusion than non-superpositional, localized excitations.} 
\label{figura1c}
\end{figure}

\section{Storing quantum information by topologically modulated
spin chains and time-inverted quantum dynamics}

To overcome this major obstacle in the construction of {\it prima
facie} working models of a quantum register we will introduce a
{\it global} (that is, site-independent) control of the total
Hamiltonian $\hat{H}_{tot}$ able to force a periodic time
reconstruction of the initial state. In the presence of a generic
time-dependent Hamiltonian, evolution of a quantum state can be
determined in the most general terms by resorting to 
the Dyson series representation. Obviously, the
structural complexity of the Dyson evolution integral
does not allow to identify all the possible Hamiltonian dynamical evolutions
yielding perfect, time-periodic quantum state reconstruction. However, the
Dyson series can be easily resummed when the Hamiltonian enjoys
the property to self-commute at different times:
$[\hat{H}(t),\hat{H}(t')]=0$ for all $t$ and $t'\neq t$. Thus, our
first prescription concerns a quantum register described by a
time-periodic modulated Hamiltonian, commuting at all times,
either when applied to all possible states, or for a subspace 
of the whole Hilbert space of states.

Under the time-commutation hypothesis, we can determine a complete
set of states $\{ \ket{\alpha} \}$ that are eigenstates of the
time-dependent Hamiltonian at all times even if the corresponding
energy eigenvalues are time-dependent functions
$\varepsilon^\alpha(t)$. Because the energy eigenstates form a
complete basis set, the initial state $\ket{\chi}$ can be written
as a linear combination: $\ket{\chi} = \sum_\alpha c_\alpha
\ket{\alpha}$, with $c_\alpha$=$\bk{\alpha}{\chi}$, and the sum
runs over the complete set of eigenstates. It is then simple to
write the evolution of the initial state after a time $t>0$,
\begin{equation}
\label{evoluzione} \ket{\psi (t)} \, = \, \sum_\alpha c_\alpha\exp
\left(-i\int_0^t \varepsilon^\alpha(\tau) d\tau
\right)\ket{\alpha} \; .
\end{equation}
From \eq{evoluzione} it is immediate to see that if at a certain
time $T>0$ all the integrals in the sum are equal or differ from
each other by integer multiples of $2 \pi$, then the initial state
$\ket{\chi}$ is perfectly reconstructed, but for an irrelevant
global phase factor. This is then the second requirement that one
needs to impose on the time-modulated dynamics in order to realize
perfect time-periodic quantum state storage.

These two basic requirements can be implemented successfully 
for spin systems with $XY$ interactions by a
scheme of quantum control based on time-inverted dynamical 
evolutions realized by suitably engineered time-dependent 
global phase factors. We then consider the following situation. 
In the presence of a simple geometry realized by placing a 
tiny solenoid in the center of a circular ring (periodic 
boundary conditions) of spins sitting at regularly spaced 
lattice sites, all the nearest-neighbor interaction amplitudes 
become complex by acquiring the same site-independent, 
global  phase $\theta$
proportional to the magnetic flux $\phi$ linked to the ring:
$\theta \propto \phi/N$, where $N$ is the total number of qubits
in the ring. The solenoid field can then be modulated in time to
achieve the control needed for perfect quantum state storage. In
the presence of a time-variable linked flux, the total Hamiltonian
\eq{GenericHamiltonian} becomes time-dependent and reads:
\begin{equation}
\label{ControlledHamiltonian} \hat{H}_{tot}(t) \, =  \, -\sum_{i}
(\lambda + \eta_i)
\left(e^{i\theta(t)}\sigma_{i}^{+}\sigma_{i+1}^{-} + H. c. \right)
 +  B\sum_{i}\sigma_{i}^{z} \; .
\end{equation}
Incidentally, we note that this scheme can apply as well
to more general systems of hopping material particles, 
thanks to the \virg{Peierls} effect, i.e. the
fact that if charged particles are in the presence of a linked
magnetic flux, the real-valued hopping amplitude between particles
(in spin language, the nearest-neighbor coupling), is transformed
in a complex-valued quantity \cite{Scalapino}.

Clearly, not all time modulations of the phase factor can realize
the desired perfect state storage. The first constraint to be
imposed is commutativity at different times:
$[\hat{H}_{tot}(t),\hat{H}_{tot}(t')]=0$. This property is verified if
and only if $\theta(t)-\theta(t')=k \pi$, with $k$ integer. This
implies that during the entire time evolution the phase must be
modulated in regular periodic jumps (steps) between two constant
values $\theta_0$ and $\theta_0+\pi$ (step-phase modulation).

\begin{figure}[t!]
\includegraphics[width=7.5cm]{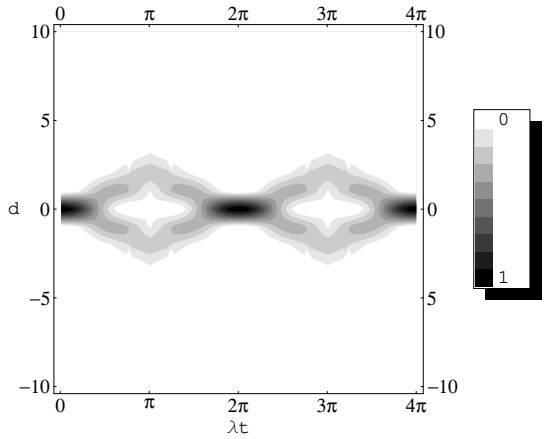}
\caption{Step-periodic time-modulation of the phase,
\eq{modulazione}: contour plot showing the evolution of the
bidimensional overlap ${\cal{F}}_{d}(t)$, for the same initial state considered
in \fig{figura1a}, as a function of the distance $d$ from site $0$
($y$-axis) and of the dimensionless time $\lambda t$ in atomic
units $\hbar=1$ ($x$-axis). Here $\lambda T =2\pi$ and
$\theta=\pi/2$. The value of the overlap increases from white
(${\cal{F}}_{d}(t) = 0$) to black (${\cal{F}}_{d}(t) = 1$).}
\label{figura2a}
\end{figure}

\begin{figure}[t!]
\includegraphics[width=7.5cm]{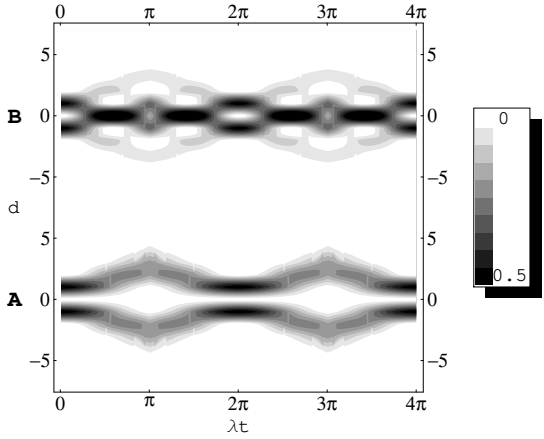}
\caption{Step-periodic time-modulation of the phase,
\eq{modulazione}: contour plot showing the evolution of the
two-dimensional overlap ${\cal{F}}_{d}(t)$, for the initial state $\ket{\chi} =
(\ket{\Psi_1} + \ket{\Psi_{-1}})/\sqrt{2}$, as a function of the
distance $d$ from site $0$ ($y$-axis) and of the dimensionless
time $\lambda t$ ($x$-axis). Here $\lambda T=2\pi$. In graph A),
$\theta$ flips periodically between $-\pi/2$ and $\pi/2$. In graph
B) it flips periodically between $0$ and $\pi$. The value of the
overlap increases from white (${\cal{F}}_{d}(t) = 0$) to black
(${\cal{F}}_{d}(t) = 0.5$).} \label{figura2b}
\end{figure}

\begin{figure}[t!]
\includegraphics[width=7.5cm]{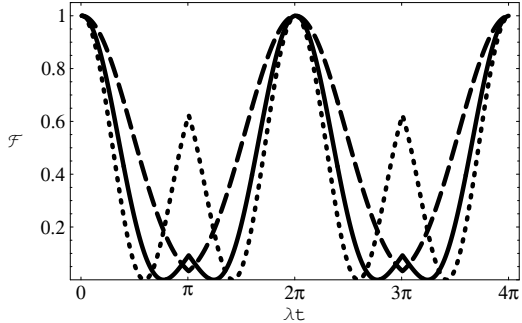}
\caption{Step-periodic time-modulation of the phase
\eq{modulazione}: One-dimensional fidelity 
${\cal{F}}_{0}(t)$ at fixed lattice site $d=0$
as a function of the dimensionless time
$\lambda t$ ($x$-axis) for different constant values $\theta_0$ 
of the phase and different initially stored states in the case of 
$\lambda T=2 \pi$. The solid line
represents the evolution of the fidelity when the initial
state is the single-magnon excitation $\ket{\Psi_0}$ 
and the phase is $\theta_0=\pi/2$. 
The dotted and the dashed lines both represent
the evolutions of the initial superposition states
$\ket{\chi} = (\ket{\Psi_1} + \ket{\Psi_{-1}})/\sqrt{2}$. The
dotted line is for the case $\theta_0=0$; the dashed line for
$\theta_0=\pi/2$. Notice that, according to the assigned value
of the phase, the speed of revival gain for the superpositional 
state can be larger or smaller than that of the nonsuperpositional
state.} 
\label{figura2c}
\end{figure}

Concerning the property that at a certain given time $T$, all the
time integrals appearing in the expression \eq{evoluzione} must be
equal (or differ by a trivial phase factor integer multiple of $2
\pi$), let us observe that, independently of the values of
$\theta$, the local term $\hat{H}_0$ (ideal register) commutes
with the $XY$ residual interaction terms in the time-dependent
Hamiltonian \eq{ControlledHamiltonian}. Hence, there exists a
complete set of eigenstates of $\hat{H}_{c}(t)$ that are as well
simultaneous eigenstates of both the local and the interaction
terms. Then, for all eigenvalues $\varepsilon^\alpha(\theta_0)$
associated to the eigenstates $\ket{\alpha}$, we have that:
$\varepsilon^\alpha(\theta_0)=\varepsilon^\alpha_{c}
(\theta_0)+\varepsilon^\alpha_{l}$, where
$\varepsilon^\alpha_{c}(\theta_0)$ and $\varepsilon^\alpha_{l}$
are, respectively, the interaction and the local contributions to
the energy. On the other hand, when $\theta$ passes from the value
$\theta_0$ to $\theta_0+\pi$, the coupling contribution to the
energy changes sign while the local one remains unchanged, so that
$\varepsilon^\alpha(\theta_0+\pi)=-\varepsilon^\alpha_{c}
(\theta)+\varepsilon^\alpha_{l}$. Therefore, any two energy
eigenvalues $\varepsilon^{\alpha}(\theta)$ and
$\varepsilon^{\alpha}(\theta+\pi)$ corresponding to the same
eigenstate $\ket{\alpha}$, differ only in the sign of the
interaction component. Then, selecting a regular step time
modulation of the phase of the form
\begin{equation}\label{modulazione}
\theta(t) \, = \, \left\{
\begin{array}{lll}
\theta \; , & \; \; \; &  0 \leq t < T/2 \; ; \\
\theta + \pi \; , & \; \; \; &  T/2 \leq t < T \; , \\
\end{array}
\right.
\end{equation}
periodically repeated for any $t \geq T$, we obtain that the
contribution of the residual interaction Hamiltonian to the
quantum state time evolution vanishes at any time $t=mT$ with $m$
arbitrary integer. Consequently, the effects of the undesired, 
residual interqubit $XY$ interactions on the quantum register are
completely eliminated, {\it even in the presence of local, static 
imperfections in the couplings}. Moreover, the result is valid
for {\it any kind} of distribution of the noise on the couplings,
be it Gaussian, or uniform, or any other probability distributution.
The effects of the storing scheme are illustrated in 
\fig{figura2a}, \fig{figura2b}, and \fig{figura2c}.

In \fig{figura2a} we again plot the time evolution of the two-dimensional
overlap ${\cal{F}}_{d}(t)$, for the same initial state considered in
\fig{figura1a}, but now under the action of Hamiltonian
$\hat{H}_{tot}(t)$ \eq{ControlledHamiltonian}. At striking
variance with the unmodulated case reported in \fig{figura1a}, the
step-phase modulated register realizes exact, time-periodic
coherent revivals of the initial state. Moreover, \fig{figura2a}
shows that the overall spatial diffusion of the state is always
confined in a well defined and extremely narrow region of the
ring. Because the topological control remains exact in the
thermodynamic limit, the number of stored states con be
arbitrarily large, possibly macroscopic. Finally, the evolution of
this kind of ``non-superpositional'' initial state is not affected
by the choice of the initial phase $\theta$. However, for the
purposes of quantum computation, the most interesting and
desirable scope is obviously the storage of superposition states.
Remarkably, if the initial state $\ket{\chi}$ is a superposition
of the form 
$$
\ket{\chi} \, = \, 
\sum_{i=1}^{M}a_{i}\ket{\Psi_{d_{i}}} \; ,
$$
i.e. such that the
excitations are distributed among different sites, the step-phase
control again yields perfect periodic state reconstruction with
extremely limited intermediate spatial spread, although some of
the details of the evolution at intermediate times can change
significantly. As an example, in \fig{figura2b} we plot
the overlap ${\cal{F}}_{d}(t)$ driven by the Hamiltonian
\eq{ControlledHamiltonian}, starting from the initial 
superposition state $\ket{\chi} =
(\ket{\Psi_1} + \ket{\Psi_{-1}})/\sqrt{2}$, and for different
values of the phase $\theta$. These results can be further
generalized to any initial state $\ket{\chi}$ with an arbitrary
number of excitations or flipped spins (arbitrarily many magnons). In this
case, one finds the same coherent time-periodic revival of the
state as in the one-magnon case, while the spatial spread becomes
a function of the size of the region along which the initial state
is extended, but remains in any case finite and limited.

\begin{figure}[t!]
\includegraphics[width=10.cm]{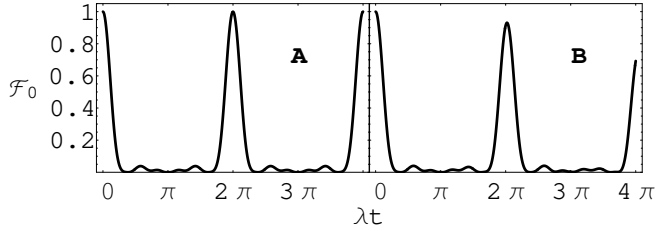}
\caption{Step-periodic time-modulation of the phase
\eq{modulazione}: One-dimensional fidelity ${\cal{F}}_{0}(t)$
at fixed lattice site $d=0$
as a function of the dimensionless time
$\lambda t$ ($x$-axis) in a ring of 90 qubits for the initial
state $\ket{\chi} = -(\sqrt{2}/3) \ket{\Psi_{20}} +
(\sqrt{1}/3) \ket{\Psi_{72}} + (\sqrt{2}/\sqrt{3})
\ket{\Psi^{(2)}_{0,5}}$ and for different values 
of the ratio $B/\lambda$. Plot A): $B/\lambda=2$, $BT=4
\pi$. Plot B): $B/\lambda=1.9$, $BT=3.8\pi$. Notice the perfect
periodic state reconstruction in plot A) (integer value of the
ratio $B/\lambda$), while a small deviation from it causes a very
slow but progressive corruption of the quantum state revivals, 
as shown in plot B). In both cases the period is fixed at the 
value $\lambda T=2 \pi$.}
\label{figura3}
\end{figure}

To better compare the different situations that can be 
realized using the step-modulated or the unmodulated phase,
we show in
\fig{figura2c} the behaviour of the one-dimensional 
fidelity at fixed lattice site $d=0$, in the presence of
periodic step modulation of the phase for different initial states
and different constant values $\theta_0$ of the phase.
It is remarkable to notice that, depending on the different
constant values of the phase that one chooses as reference values
to operate with the control scheme, the approach to the full revival 
for the superpositional states can be larger or smaller than that of 
the non-superpositional states.
An important point is to identify all the possible 
classes of states that can be perfectly stored using our 
method. Obviously, the answer to
this question involves not only the interaction Hamiltonian,
but the local, diagonal term $\hat{H}_0$ as well. 
If the period $T$ of the step-phase control is gauged so
that 
$$
BT \, = \, 2l\pi 
$$ 
with $l$ integer, then {\it any initial quantum
state} is reconstructed {\it exactly} at all times $t$ integer
multiples of $T$. If instead $BT \neq 2l\pi$, perfect quantum
state storage is still achieved in the subspace of all states that
are linear combinations of local states with equal
magnetization. To gain further understanding of this result, let
us consider generalizations of the locally excited states $\ket{\Psi_d}$
by introducing the n-times locally excited states
\begin{equation}
\ket{\Psi^{(n)}_{d_1,...,d_n}} \, = \, \prod_{j=d_1,...,d_n}
\ket{\uparrow_j}\prod_{i\neq d_1,...,d_n }\ket{\downarrow_i} \; . 
\label{generalizedmagnons}
\end{equation}
We thus wish to analyze the case in which the state to be stored is a
superposition of local states with different values of the 
magnetization, and either integer or real values of the ratio
$B/\lambda$. To this end, in \fig{figura3} we plot the one-dimensional 
fidelity  ${\cal{F}}_{0}(t)$ at fixed lattice site $d=0$ for an initially
stored state of the form
$$
\ket{\chi} \, = \, -(\sqrt{2}/3) \ket{\Psi_{20}} \, + \, 
(\sqrt{1}/3) \ket{\Psi_{72}} \, + \, (\sqrt{2}/\sqrt{3})
\ket{\Psi^{(2)}_{0,5}} \; ,
$$
where the first two terms in the superposition are one-magnon
states with flipped spin, respectively, at site $d=20$ and
$d=72$, and the last term is the superposition of two one-magnon
states with the excitation placed, respectively, on site $d=0$ and
site $d=5$. The initial state is thus chosen to be a superposition
of states belonging to different subspaces of Hilbert space.
In the present case, the
subspace of single excitations and the subspace of states of
the form Eq. (\ref{generalizedmagnons}). As already discussed,
such a generic initial state achieves perfect periodic reconstruction
in the case of an integer value of the ratio $B/\lambda$. In the case
of noninteger ratios $B/\lambda$, the state undergoes quasi-perfect
periodic revivals that after few cycles begin to deteriorate. Adding
more terms to the initial superposition greatly increases the number
of cycles with quasi-perfect reconstruction. For this kind of initial
states the analysis cannot be performed analytically in all details, 
as previously done for states belonging to the same subspace. However,
numerics can be easily implemented for spin rings of finite, but large
size. In the instance considered, we solve numerically the unitary
evolution of ${\cal{F}}_{0}(t)$ for a closed chain of 90 sites. Considering
larger rings yields qualitatively identical results.

\section{Other sources of noise}

\begin{figure}[h!]
\includegraphics[width=8.cm]{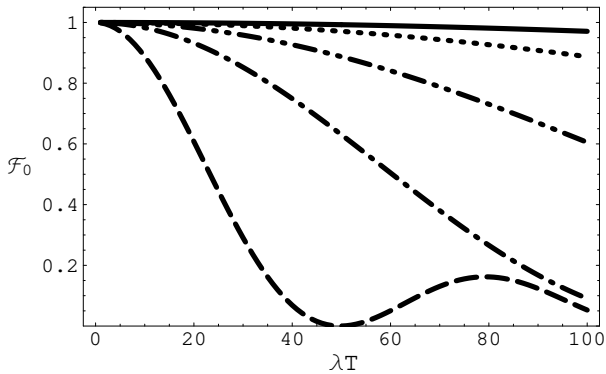}
\caption{Fidelity ${\cal{F}}_{0}(t)$ on site $i=0$ as a function
of the number of periods $\lambda T$ for the initial one-magnon
state $\ket{\Psi_0}$ when the step-periodic time-modulated phase
$\theta(t)$ is replaced with its finite-harmonic Fourier
approximations, in increasing order. Dashed line: first 5
harmonics; dot-dashed line: first 13; dot-dot-dashed line: first
25; dotted line: first 50; solid line: first 100 harmonics.}
\label{figura3a}
\end{figure}

\begin{figure}[h!]
\includegraphics[width=8.cm]{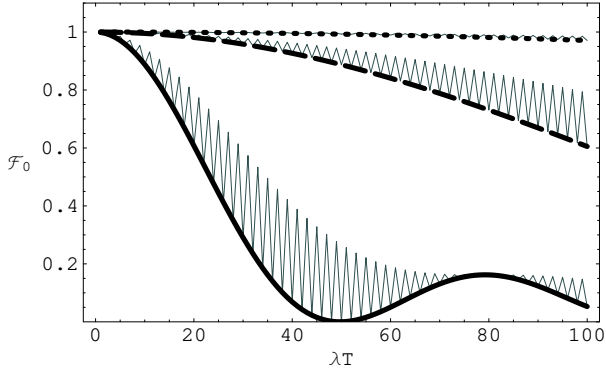}
\caption{Fidelity ${\cal{F}}_{0}(t)$ on site $i=0$ as a function
of the number of periods $\lambda T$ for the initial one-magnon
state $\ket{\Psi_0}$ (black curves) versus the fidelity of the initial state
$\ket{\chi} = (\ket{\Psi_1} + \ket{\Psi_{0}})/\sqrt{2}$ (grey curves),
when the step-periodic time-modulated phase $\theta(t)$ is replaced with
its finite-harmonic Fourier approximations, in increasing order.
Solid line: first 5 harmonics; dashed line: first 25; dotted line:
first 100 harmonics.} 
\label{figura3b}
\end{figure}

Concerning the issue of practical implementations, it is crucial
to verify that the storing scheme does not depend critically on a
perfect realization of the step-phase modulation. To analyze this
dependence, in \fig{figura3a} we show the evolution of the fidelity
${\cal{F}}_{0}(t) = |\langle \Psi_0|\psi(t)\rangle|^2$ for the
same initial one-magnon state $\ket{\Psi_0}$ previously studied, as a
function of the number of periods, when the step-periodic,
time-modulated phase $\theta(t)$ \eq{modulazione} is approximated
by its Fourier decompositions, truncated at various finite orders.
Remarkably, even when considering only the first $100$ harmonics,
the fidelity remains close to the ideal limit ${\cal{F}}_{0}(t)=1$
for very long times. In \fig{figura3b} we compare this behaviour
with that of a superpositional initial state. Remarkably, the fidelity
of initial superpositions, although wildly oscillatory, is always larger
than that of non-superpositional initial states.
These results prove the stability of the
finite-harmonic approximation and, as a consequence, the
robustness of the storing protocol against imperfections in the
external control of the phase.
Another point that needs to be addressed is the effect on the 
step-phase control of static, site-dependent imperfections in 
the magnetic field $B$ that enters in the Hamiltonian $\hat{H}_0$. 
Remarkably, it turns out that the global topological
control on the off-diagonal Hamiltonian terms allows partial
control on the diagonal terms as well. Comparing the time
evolutions with and without step-phase modulation, in the presence
of imperfections $\delta_i$ in $B$, with Gaussian distribution of
not too broad half-width $\sigma_{\delta}$, one finds that in the
latter case quantum diffusion grows indefinitely (as expected),
while in the former case it remains limited and the fidelity
undergoes revivals that can reach unity in specific cases.
Moreover, the effects of the $XY$ residual interactions continue
to be completely suppressed \cite{IJQC}.

\section{Conclusions and outlook}

In Conclusion, we have introduced a scheme for the storage of
information in a quantum register by a global, topological quantum
control that realizes periodic, perfect state reconstruction in
periodic qubit chains (rings). This kind of quantum register
controlled by step-phase modulations is able to fully cancel the
effects of environmental (residual) interactions of the $XY$ type,
even in the presence of local, static imperfections in the inter-qubit 
couplings. Moreover, the scheme is robust even in the presence of other
sources of noise, such as phase modulations of finite precision
and local static noise on the computational Hamiltonian
\cite{IJQC}. The study of the effects of dynamic imperfections is
under way \cite{futuro}. However, it is already clear at this
stage that, at least either in the ultrafast or in the adiabatic limit, the
present storage scheme remains unaffected due to its global,
topological nature.

\vspace{0.2cm}

We thank Simone Montangero and Giuseppe Falci for stimulating discussions.
Financial support from INFM, INFN and MIUR is acknowledged.


\vspace{0.6cm}

\begin{thebibliography}{99}

\bibitem{Bose} S. Bose, Phys. Rev. Lett. {\bf 91}, 207901 (2003).

\bibitem{Depasquale} F. de Pasquale, G. Giorgi, and S. Paganelli,
Phys. Rev. Lett. {\bf 93}, 120502 (2004).

\bibitem{Subramanyam} V. Subrahmanyam, Phys. Rev. A {\bf 69}, 034304 (2004).

\bibitem{Datta}  M. Christandl, N. Datta, A. Ekert, and A. J. Landahl,
Phys. Rev. Lett. {\bf 92}, 187902 (2004).

\bibitem{Song} Z. Song and C. P. Sun, quant-ph/0412183, and references therein.

\bibitem{Lukin} M. D. Lukin, Rev. Mod. Phys. {\bf 75}, 457 (2003).

\bibitem{Lukin2} M. Fleischhauer and M. D. Lukin, Phys. Rev. Lett. {\bf 84}, 5094
(2000); Phys. Rev. A {\bf 65}, 022314 (2002).

\bibitem{Lukin3} J. M. Taylor, C. M. Marcus, and M. D. Lukin, Phys. Rev. Lett.
{\bf 90}, 206803 (2003).

\bibitem{nano1} E. Pazy, I. D'Amico, P. Zanardi, and F. Rossi, Phys. Rev. B {\bf 64},
195320 (2001).

\bibitem{nano2} Y. Li, P. Zhang, P. Zanardi, and C. P. Sun, Phys. Rev. A {\bf 70},
032330 (2004).

\bibitem{Zoller} A. Imamoglu, E. Knill, L. Tian, and P. Zoller, Phys. Rev. Lett.
{\bf 91}, 017402 (2003).

\bibitem{Poggio} M. Poggio, G. M. Steeves, R. C. Myers, Y. Kato, A. C. Gossard, and D. D. Awschalom,
Phys. Rev. Lett. {\bf 91}, 207602 (2003).

\bibitem{Gingrich}  R. M. Gingrich, P. Kok, H. Lee, F. Vatan, and J. P. Dowling,
Phys. Rev. Lett. {\bf 91}, 217901 (2003).

\bibitem{Kitaev} E. Dennis, A. Kitaev, A. Landahl, and J. Preskill, J. Math. Phys. {\bf 43}, 4452 (2002).

\bibitem{Georgeot} B. Georgeot and D. L. Shepelyansky, Phys. Rev. E {\bf 62}, 3504 (2000);
Phys. Rev. E {\bf 62}, 6366 (2000).

\bibitem{Roadmap} See the progress updates at the quantum computation roadmap
website http:$//$qist.lanl.gov$/$qcomp${}_{}$map.shtml and references therein.

\bibitem{Divincenzo} D. P. DiVincenzo, Science {\bf 270}, 255 (1995).

\bibitem{Montangero}  G. De Chiara, D. Rossini, S. Montangero, and
R. Fazio, Phys. Rev. A {\bf 72}, 012323 (2005).

\bibitem{Schoen} Y. Makhlin, G. Sch\"on and A. Shnirman, Rev. Mod. Phys. {\bf 73}, 357
(2001).

\bibitem{ControlJOB} S. M. Giampaolo, F. Illuminati, and G. Mazzarella, 
J. Opt. B: Quantum Semiclassical Opt. {\bf 7}, S337 (2005).

\bibitem{Fazio} L. Amico, A. Osterloh, F. Plastina, R. Fazio,
and G. M. Palma, Phys. Rev. A {\bf 69}, 022304 (2004).

\bibitem{Scalapino} D. J. Scalapino, S. R. White, and S. Zhang,
Phys. Rev. B {\bf 47}, 7995 (1993).

\bibitem{IJQC} S. M. Giampaolo, F. Illuminati, A. Di Lisi, and G. Mazzarella, 
Int. J. Quant. Inf. (to be published), and quant-ph/0506227.

\bibitem{futuro} S. M. Giampaolo, F. Illuminati, and S. Montangero, in preparation.

\end{thebibliography}
\end{document}